\documentclass{iaus}
\usepackage{graphicx}

\newcommand{\mnras}{MNRAS}
\newcommand{\apss}{ApSS}
\newcommand{\aap}{A\&A}
\newcommand{\apj}{ApJ}
\newcommand{\araa}{ARAA}
\newcommand{\aj}{AJ}
\newcommand{\pasj}{PASJ}
\newcommand{\apjl}{ApJ}
\newcommand{\apjs}{ApJS}
\newcommand{\pasp}{PASP}

\title[The Gould's Belt Distances Survey] 
{The Gould's Belt Distances Survey}

\author[Laurent Loinard]   
{Laurent Loinard}

\affiliation{Centro de Radioastronom\'{\i}a y Astrof\'{\i}sica, Universidad Nacional Aut\'onoma de M\'exico,  Apartado Postal 3-72 (Xangari), 58089 Morelia, Michoac\'an, M\'exico\\ email: {\tt l.loinard@crya.unam.mx} \\[\affilskip]}

\pubyear{2012}
\volume{289}  
\pagerange{1--8}
\jname{Advancing the Physics of Cosmic Distances}
\editors{R.\ de Grijs and G.\ Bono, eds.}
\begin{document}

\maketitle

\begin{abstract}
Very Long Baseline Interferometry (VLBI) observations can provide the position
of compact radio sources with an accuracy of order 50 micro-arcseconds. 
This is sufficient to measure the trigonometric parallax and proper motions 
of any object within 500 pc of the Sun to better than a few percent. Because they are
magnetically active, young stars are often associated with compact radio emission
detectable using VLBI techniques. Here we will show how VLBI observations have already 
constrained the distance to the most often studied nearby regions of star-formation (Taurus,
Ophiuchus, Orion, etc.) and have started to provide information on their internal structure and 
kinematics. We will then briefly describe a large project (called {\it The Gould's Belt Distances 
Survey}) designed to provide a detailed view of star-formation in the Solar neighborhood 
using VLBI observations.
\end{abstract}

\firstsection 
\section{Introduction}

Gould's Belt (See Poppel 1997 for a recent review) is a prominent local Galactic structure that contains most 
nearby O and B stars, as well as about 2 million solar masses of interstellar material. It is a flattened system 
of radius $\sim$ 1 kpc, inclined by roughly 18$^\circ$ relative to the Galactic plane, and centered at about 100 
pc from the Sun in the direction $\ell$ $\sim$ 180$^\circ$. It is named after the American astronomer Benjamin 
Gould (also notable for being the first american to receive a Ph.D.\ in astronomy, and the founder of the Astronomical 
Journal) who described it in 1879. It is worth pointing out, however, that Sir John Herschel had already 
identified many of its salient features in the late 1840's. The stars in Gould's Belt exhibit a peculiar 
kinematics clearly indicative of non-uniform expansion. Its age is estimated to be of order 30 Myr, 
but its origin remains very unclear. An impulsive event, such as the impact of a dark matter halo cloud 
similar to Smith's cloud (see Lockman et al.\ 2008) with a large molecular cloud in the Galactic disk is 
a plausible suggestion. The complex kinematics of the structure, however, suggests that local events within 
the structure (e.g.\ supernovae explosions) also contribute to its overall morphology. 

Gould's Belt contains essentially all the sites of on-going star-formation within 500 pc -- Ophiuchus, Lupus, 
Taurus, Orion, the Aquila Rift and Serpens region, etc. The detailed study of these regions has played,
and continues to play, a central role in the development of our understanding of star-formation in general. 
This can be seen, for instance, from the fact that in the last few years alone, several large space programs like 
the Spitzer c2d survey (Evans et al.\ 2009), the XMM Newton Extended Survey of Taurus (G\"udel et al.\ 2008), or 
the Herschel Gould Belt Survey (Andr\'e et al.\ 2010) have focused on these regions. Ancillary ground-based 
observations such as the JCMT Gould's Belt Survey (Ward-Thompson et al.\ 2007) have also been gathered. 
The analysis of this wealth of information has been somewhat affected by the lack of accurate distances to the 
regions within Gould's Belt, which are rarely better than 30\% and often significantly worse. This problem is 
compounded by the fact that several of these regions (as we shall see, the Aquila/Serpens region is a case in 
point) might contain structure at different distances along the line of sight.

The {\it Gould's Belt Distances Survey} is a large project aimed at measuring the astrometric elements
of about 200 low-mass young stars distributed across five of the most prominent star-forming regions in
Gould's Belt (Ophiuchus, Taurus, Perseus, Aquila/Serpens, and Orion) using Very Long Baseline Interferometry 
(VLBI) multi-epoch radio observations. The immediate goal of these observations is to accurately measure the 
distance, three-dimensional structure, and internal kinematics of these regions. This will have important implications both 
for the study of star-formation in general, and for the analysis of the very origin of Gould's Belt. In this contribution, 
we will first present the technique used and a handful of results obtained towards nearby star-forming regions.
We then proceed to describe the Gould's Belt Distances Survey itself, and the current status of the project.

\section{VLBI astrometry of low-mass young stars}

Very Long Baseline Interferometry (VLBI; e.g.\ Thompson, Moran \& Swenson 2001) is an 
observing technique that can readily provide astrometric accuracies of order 30--60 $\mu$as, 
for compact radio sources of sufficient brightness ($T_b$ $\approx$ 10$^7$ K). Such 
sources exist in all star-forming regions, because low-mass young stars are often magnetically 
active. The gyration of relativistic electrons in the strong magnetic fields surrounding low-mass 
young stars produce radio emission detectable with VLBI instruments (Dulk 1985). This 
emission is normally confined to regions extending only a few stellar radii around the 
stars, and therefore remain very compact even in the nearest star-forming regions 
(at $d$ $\approx$ 100 pc). In high mass star-forming regions, water and methanol masers 
(which are also detectable with VLBI instruments) are common, and offer an interesting 
alternative to magnetically active low-mass stars. Here, however, we will be primarily 
concerned with low and intermediate mass star-forming regions where masers are of limited 
use because they are rare and often strongly variable. 

\begin{figure}[b]
\begin{center}
 \includegraphics[width=3.0in,angle=-90]{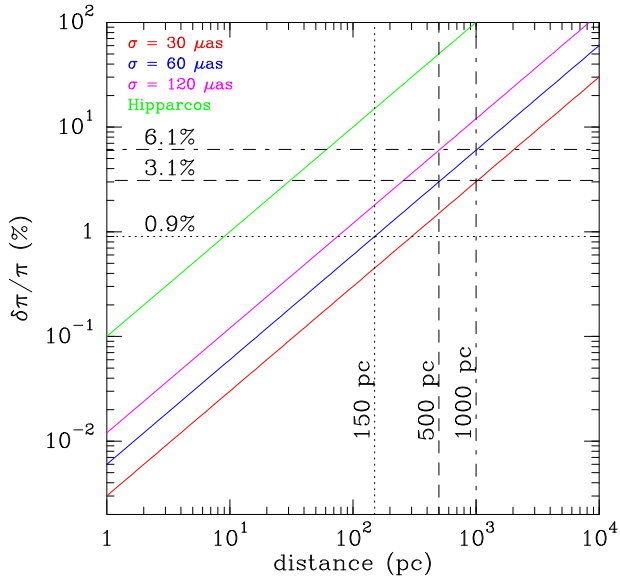} 
 \caption{Relative parallax uncertainty as a function of distance for typical VLBI
 astrometric accuracies (30, 60, 120 $\mu$as). For reference, the mean Hipparcos
 uncertainty (1 mas) is also shown. Representative distances (150, 500, and 1000 pc)
 are indicated.}
   \label{fig1}
\end{center}
\end{figure}

The implication of the astrometric accuracy reached in VLBI experiment for distance
measurements is shown in Figure 1, where we consider the relative parallax error
($\delta\pi / \pi$) as a function of distance for astrometric accuracies typical of VLBI
observations (30, 60, and 120 $\mu$as). For reference, the mean accuracy of Hipparcos
(1 mas) is also shown. It can be seen that for any source within about 1 kpc, VLBI 
observations can yield accuracies of a few percent on the distance measurement.
To demonstrate that this can be achieved in practice, we show in Figure 2 the trajectory 
described on the plane of the sky by T Tau Sb, one of the southern companions of the 
famous object T Tauri, as characterized by multi-epoch VLBI observations (see Loinard 
et al.\ 2007 for details). That trajectory is the combination of an elliptic parallax component 
and the proper motion of the source --oriented towards the south-east, in this specific 
case. Note that the entire size of the figure is only about 10 mas. The distance obtained
from these observations (146.7 $\pm$ 0.6 pc), has an uncertainty of 0.4\%.

We now proceed to describe the results that have been obtained so far for nearby 
star-forming regions using multi-epoch VLBI observations. 

\begin{figure}[b]
\begin{center}
 \includegraphics[height=5.0in,angle=-90]{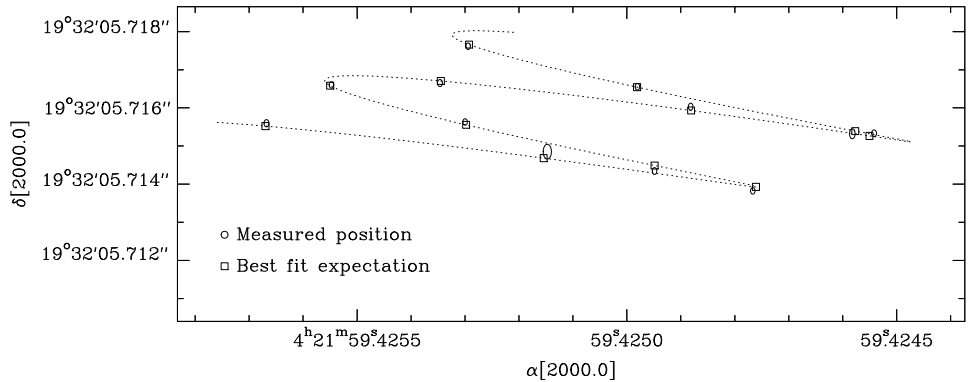} 
 \caption{Trajectory of T Tau Sb on the plane of the sky as measured by
multi-epoch observations. The data were collected at 12 epochs from September
2003 to July 2005, and are shown as ellipses whose sizes represent the error
bars. The dotted curve shows the best fit with a combination of parallax and
proper motion. The squares are the positions predicted by the best fit at the
twelve epochs (see Loinard et al.\ 2007 for details).}
   \label{fig1}
\end{center}
\end{figure}

\subsection{Ophiuchus}

During many years, Ophiuchus was thought to be at 165 pc (Chini 1981). More
recently, however, shorter distances have been preferred. For example, de Geus 
et al.\ (1989) found a mean photometric distance of 125 $\pm$ 25 pc for the stellar
population associated with Ophiuchus. Knude \& H\"og (1998), who examined the 
reddening of stars in the direction of Ophiuchus as a 
function of their Hipparcos distances, found a clear extinction jump at 120 pc and 
agued that this corresponds to the distance of the main dust component in Ophiuchus. 
Using a similar method, Lombardi et al. (2008) also reported a distance of about 120 
pc for the Ophiuchus core. 

Using VLBI observations, Loinard et al.\ (2008) measured the trigonometric parallax
of two stars in the Ophiuchus core (S1 and DoAr21), and found a mean value of 
120.0 $\pm$ 4.5 pc. The Ophiuchus core is only about 2 degrees across, 
corresponding to a linear radius of about 2 pc. The depth of the region is expected 
to be similar, so systematic errors on the distance to individual objects in the Ophiuchus 
core due to its depth are expected to be of that order. To account for that effect, we
add quadratically 2 pc to the formal errors reported by Loinard et al.\ (2008), and
conclude that the distance to the Ophiuchus core should be assumed to be 120 $\pm$ 
5 pc.

In addition to its core, Ophiuchus contains several filamentary structures known 
as the {\it streamers} (see Wikling et al. 2008). The streamers correspond to 
prominent dust clouds (such as Lynds 1689, 1709, 1712, or 1729) and extend to 
about 10 pc from the core. They are thought to be physically associated with the
Ophiuchus core, and should therefore be at similar distances (within about 10 pc,
their physical extent on the plane of the sky). There are several indications that this 
might indeed be the case. For instance, Lombardi et al.\ (2008) found some indications 
that the streamers might be 5 to 10 pc nearer than the core. Similarly,  Schaefer et al.\ 
(2008) found a distance (based on orbit modeling) to the binary young stellar object 
Haro 1--14c, located in the north-eastern streamer, of 111$^{+19}_{-18}$ pc.

A somewhat puzzling result in that respect comes from the determination by Imai 
et al.\ (2007) of the trigonometric parallax of the very young stellar system 
IRAS~16293--2422 in the eastern streamer L1689. Based on VLBI observations of
water masers associated with IRAS~16293--2422, they determine a distance 
of 178$^{+18}_{-37}$ pc. Since the entire Ophiuchus complex is only about 10 pc
across on the plane of the sky, it is very unlikely to be 60 pc deep. Thus, if the
distance determination of Imai et al.\ (2007) were confirmed, it would indicate that
at least two unrelated star-forming regions coexist along the line of sight toward
Ophiuchus. One region, associated with the core and north-eastern streamers, 
would be at 120 pc, while the region associated with the eastern cloud L1689
would be several tens of pc farther. It would clearly be very important to obtain an 
independent confirmation of the distance to L1689.

\subsection{Taurus}

Taurus is perhaps the region that has been most instrumental to the development 
of our understanding of star-formation (see Kenyon et al.\ 2008 for a recent review).
Its mean distance has long been known to be about 140 pc (Kenyon et al.\ 1994),
and the total extent of the region on the plane of the sky is about 10 degrees (or 25
pc). Roughly speaking, Taurus is composed of three parallel filaments 
each about 2 degrees thick. Since those filaments have no reason to be 
orientated perpendicularly to the line of sight, it is quite conceivable that  the near
side of each filament might be about 25 pc closer than its far side. 

In a series of papers (Loinard et al.\ 2005, 2007; Torres et al.\ 2007, 2009, 2012), 
we have reported measurements of the trigonometric parallax of five young stars 
distributed across the Taurus complex. Three of these stars (Hubble 4, HDE~283572, 
and V773 Tau) are located toward the same portion of Taurus, associated 
with the prominent dark cloud L1495. Interestingly, all three stars are found to be
at a similar distance (132.8 $\pm$ 0.5 pc, 128.5 $\pm$ 0.6 pc, and 132.9 $\pm$
2.4 pc). They also appear to share the same kinematics as measured by their 
proper motions and radial velocities, and therefore appear to belong to a coherent 
spatio-kinematical structure. The weighted mean of the three distance measurements
(131.4 pc) clearly provides a good estimate of the distance to that specific portion of
Taurus. The dispersion about that mean (2.4 pc), on the other hand, must provide
of good estimate of the local depth of the Taurus complex. Interestingly, L1495
and its associated stellar population is about 2 degrees across. This corresponds to 
a physical radius of about 2.5 pc for that region; the stellar population associated with
L1495 appears to be about as deep as it is wide.

The other two stars for which VLBI trigonometric parallaxes have been measured 
are T Tauri Sb and HP Tau/G3. They are located, respectively, to the south-east and
north-east of the Taurus complex. T Tauri was found to be at 146.7 $\pm$
0.6 pc, while HP Tau is at 161.9 $\pm$ 0.9 pc (Loinard et al.\ 2007, Torres et al.\ 
2009). Their proper motions are similar to one another, but significantly different 
from those of the stars associated with L1495.  The errors on our distance 
measurements (a few pc) are about 10 times smaller than the separation (about 
30 pc) between the nearest and farthest stars in our sample. Thus, we are clearly
resolving the depth of the Taurus complex along the line-of-sight.

Additional accurate distance measurements will be needed to reach definite 
conclusions regarding the three-dimensional structure of Taurus, but a few
preliminary conclusions can be made. Taurus clearly appears to be about as 
deep as it is wide (25--30 pc), with the near edge associated with the western 
side and the far edge corresponding to the eastern side. In that configuration,
the filaments in Taurus would be oriented almost along the Galactic center--
Galactic anticenter axis (Ballesteros-Paredes et al.\ 2009). The space velocity
of stars near the western edge of the complex are significantly different from that
of the stars near the eastern edge. 

Clearly, observations similar to those described here for several tens of young
stars distributed across the entire complex would have the potential to reveal both
the three-dimensional structure and the internal kinematics of the stellar population
associated with Taurus. This would have important consequences for our 
understanding of star-formation.

\subsection{Perseus}

The Perseus cloud (Bally et al.\ 2008) contains a  number of important sub-regions
(for instance NGC1333 and IC 348; Walawender et al.\ 2008, Herbst 2008) where
star-formation is particularly active. For a long time, the distance to Perseus as a
whole was assumed to be 350 pc, following Herbig \& Jones (1983). Specifically,
this distance was for NGC 1333, and was obtained from an analysis of several 
previous measurements. Somewhat later, Cernis et al.\ (1990) argued for a 
shorter distance to NGC~1333 (220 pc) based on extinction measurements. Many 
authors, however, have continued to use values between 300 and 350 pc. Indeed,
it is not entirely clear if the different sub-regions are physically associated, and
even if they are, there could be a significant distance spread because the entire 
region is about 6 degrees across (corresponding to 25--30 pc).

Hirota et al.\ (2008, 2011) have measured the distance to two young stellar objects 
in Perseus using VLBI observations of their associated water masers. They obtain
a distance of 235 $\pm$ 18 pc for the source SVS 13 in NGC~1333, and 232 $\pm$
18 pc for a young stellar source associated with the dark cloud L 1448 at the western
edge of the Perseus complex. The projected separation between the two sources
considered by Hirota et al.\ (2008, 2011) is about 1.5 degrees, or 6 pc. Their
result that the two sources are at similar distances indicates that the western portion
of Perseus can be assumed to be located at about 230--235 pc from the Sun. VLBI
determination of the distance to young stars in the central part of Perseus (associated
with B1) and of the eastern portions (B5 and IC 348) would enable a complete
description of the structure of that important region.

\subsection{Serpens and Aquila}

The Serpens molecular cloud complex (Eiroa et al.\ 2008) is an important region of 
star-formation extending for several degrees on the plane of the sky. Its distance has been a matter of 
some controversy over the years, with estimates ranging from 250 pc to 650 pc (see
Eiroa et al.\ 2008 and Dzib et al.\ 2010 for a discussion). The distance used by most
authors in the last 15 years is that reported by Straizys et al.\ (1996) based on stellar
photometry: $d$ = 259 $\pm$ 37 pc. More recently, Straizys et al.\ (2003) extended
their work to include many more stars covering an area of about 50 square degrees, 
and concluded that the front edge of the cloud was at 255 $\pm$ 55 pc, and its depth
about 80 pc. In projection, Serpens is located toward a larger complex of molecular 
clouds called the Aquila Rift (Dame \& Thaddeus 1985; Prato et al.\ 2008) believed to
be at a distance of about 200 pc. The similarity between the distance to the Serpens
cloud found by Straizys and coworkers, and that of the Aquila Rift suggests a physical
association between the two.

Using VLBI observations, Dzib et al.\ (2010) measured the trigonometric parallax
of the binary system EC95 in the Serpens core region. They obtain a distance of
414.9 $\pm$ 4.4 pc, significantly larger than the value usually adopted for Serpens.
The Serpens core is a dense sub-region of the Serpens cloud, with an extent
of only about 5$'$ (0.6 pc) on the plane of the sky. As a consequence, the contribution
of depth to the uncertainty of the distance to the Serpens core is expected to be very limited,
and that distance can be assumed to be 415 $\pm$ 5 pc. The
Serpens cloud, on the other hand is about 3 degrees across, or 25 in diameter.
The distance to the cloud as a whole should therefore be taken to be 415 $\pm$ 
25 pc. 

A detailed analysis of the possible reasons for the discrepancy between the
most popular distance used for Serpens in the last 15 years, and that obtained
directly from a trigonometric parallax is given in Dzib et al.\ (2010). Their 
conclusion is that the method used by Straizys et al.\ (1996, 2003) would
naturally be sensitive to the first dust contribution along the line of sight, 
which might be associated with the clouds in the Aquila Rift rather than
those of Serpens. In that interpretation, the Serpens and Aquila Rift
clouds would not be physically associated, but merely located along the
same line of sight. This is certainly not unexpected since that line of sight 
is roughly in the direction of the Galactic plane. It will be very important to
obtain VLBI parallax measurements of other stars in the direction of Serpens
to confirm the distance measurement of Dzib et al.\ (2010) and clarify the
relation between Serpens and the Aquila Rift.

\subsection{Orion}

Together with Taurus, Orion is undoubtedly the most often studied region of
star-formation. The efforts to measure the distance to Orion have largely been
concentrated on the Orion Nebula region (an interesting and entertaining account
of the history of the distance to Orion is given by Muench et al.\ 2008). For 
decades, the accepted distance was 480 pc following Genzel et al.\ (1981). 
In the last five years, however, several VLBI parallax measurements have been 
published. Hirota et al.\ (2007)
obtained a distance of 437 $\pm$ 19 pc using VLBI observations of water masers
in the BN/KL region, whereas Sandstrom et al.\ (2007) found 389 $\pm$ 23 pc
from VLBI continuum observations of a flaring star. The issue was settled by
Menten et al.\ (2007) and Kim et al.\ (2008) who independently obtained highly 
consistent measurements (414 $\pm$ 7 pc and 418 $\pm$ 6 pc, respectively). 

It should noted, however, that the distances quoted so far are only for the
Orion Nebula region. The Orion cloud complex is about 100 pc across, so
different parts of the complex are most certainly at different distances. In
addition, there are a number of dark clouds (e.g.\ L1617 and 
L1622; Reipurth et al.\ 2008) located in the vicinity of the Orion complex,
but whose relation to the Orion clouds themselves is unclear. There are
some indications that some of these clouds might be significantly closer
than the Orion Nebula. Thus, a comprehensive VLBI program aimed at 
establishing the distance to the different parts of the Orion and surrounding 
clouds appears to be necessary.

\section{The Gould's Belt Distance Survey}

The VLBI observations described so far have targeted a total of only about a
dozen different objects, but have already significantly refined our knowledge 
of the distance to the nearest regions of star-formation. In several cases, they
have provided the first direct indications of the three-dimensional structure of 
these regions. 

One of the world's premier VLBI instruments (the Very Long Baseline Array
--VLBA) has just undergone a major upgrade, which increased its 
bandwidth (and therefore its sensitivity) by a factor of several. Thanks to
that upgrade, observations similar to those presented here will become
feasible for several hundred young stars. Taking advantage of that possibility,
we have initiated a large project (the {\it Gould's Belt Distances Survey}) aimed
at measuring the distance to about 200 young stars distributed across the five
regions described in Section 2 (Ophiuchus, Taurus, Perseus, Serpens and
Orion). The first stage of the project (which is now completed) used about 120 hours of observing time
on the Expanded Very Large Array (EVLA) to identify adequate targets. 
Following that first stage, a total of about 2000 hours of VLBA time will be 
dedicated to the astrometric observations themselves. 
The VLBA observations have just started, and will last for a total of about 4 years. 

The final goal of the {\it Gould's Belt Distance Survey} is to estimate with 
unprecedented accuracy the mean distance, three-dimensional structure 
and internal kinematics of the five regions described in Section 2. This
will have important consequences both for the study of star-formation,
and for our understanding of the local structure of the Milky Way. In
particular, our observations will shed new light on the very origin of Gould's 
Belt. The results will be posted as they are obtained on the project's web
site: http://www.crya.unam.mx/$\sim$l.loinard/Gould/.

\acknowledgements 
L.L. acknowledges the support of the von Humboldt Stiftung, DGAPA, UNAM and 
CONACyT, M\'exico. The National Radio Astronomy Observatory is a facility of the 
National Science Foundation operated under cooperative agreement by Associated Universities, Inc.

\end{document}